\documentclass{copernicus}

 \hack{\hyphenation{mo-le-cu-lar}}
 \hack{\hyphenation{pa-ra-me-ters}}
 \hack{\hyphenation{transport}}
 \hack{\hyphenation{consideration}}
 \hack{\hyphenation{appro-xi-ma-tion}}
 \hack{\hyphenation{Fur-ther-more}}

\begin{document}

\title{Ionisation as indicator for cosmic ray acceleration}

\author[1]{F.~Schuppan}
\author[2]{C.~R\"{o}ken}
\author[1]{N.~Fedrau}
\author[1]{J.~Becker Tjus}

\affil[1]{Ruhr-Universit\"{a}t Bochum, Fakult\"{a}t f\"{u}r Physik \& Astronomie,
Institut f\"{u}r Theoretische Physik IV, 44780~Bochum, Germany}
\affil[2]{Universit\"{a}t Regensburg, Fakult\"{a}t f\"{u}r Mathematik, 93040 Regensburg, Germany}

\correspondence{F.~Schuppan (florian.schuppan@rub.de) \\ \vspace{10pt} Received: 28 February 2014 -- 
Revised: 16 April 2014 -- Accepted: 9 May 2014 -- Published: 2 June 2014}

\runningtitle{Ionisation as indicator for cosmic ray acceleration}
\runningauthor{F.~Schuppan et al.}

\received{28 February 2014}
\revised{16 April 2014}
\accepted{9 May 2014}
\published{2 June 2014}

\firstpage{1}

\maketitle  

\begin{abstract}
Astrospheres and wind bubbles of massive stars are believed to be sources of
cosmic rays with energies $E\lesssim 1$\,\unit{TeV}. These particles are not
directly detectable, but their impact on surrounding matter, in particular
ionisation of atomic and molecular hydrogen, can lead to observable
signatures. A correlation study of both gamma ray emission, induced by
proton-proton interactions of cosmic ray protons with kinetic energies
$E_\mathrm{p}\ge 280$\,\unit{MeV} with ambient hydrogen, and ionisation
induced by cosmic ray protons of kinetic energies $E_\mathrm{p}<
280$\,\unit{MeV} can be performed in order to study potential sources of
(sub)TeV cosmic rays.
\end{abstract}

\introduction While supernova remnants are the main candidate for the
acceleration of Galactic cosmic rays \citep[see,
e.g.,][]{baade1934,ackermann2013}, i.e., cosmic rays with energies $E\lesssim
10^{18}$\,\unit{eV}, there are other acceleration sources contributing to the
total diffuse flux of cosmic rays. Among these are astrospheres and wind
bubbles of massive stars, which are capable of accelerating cosmic rays up to
kinetic energies of $E\lesssim 1$\,\unit{TeV} and may, in fact, be dominant
in this energy domain \citep[see,
e.g.,][]{casse80,voelk82,binns05,scherer08}. The cosmic ray flux from a
single supernova remnant vastly exceeds that from a single astrosphere or
wind bubble. However, despite the lower amount of cosmic rays from
astrospheres and wind bubbles, they can be measured indirectly via
proton-proton interactions of GeV cosmic ray protons with ambient hydrogen,
leading to the formation of neutral pions which, in turn, decay into two
gamma ray photons. Formed in sufficiently large numbers, these gamma rays can
be detected with instruments such as FermiLAT, H.E.S.S., and CTA. When cosmic
rays are accelerated to kinetic energies of several GeV, there are also
particles of lower energy to be expected. Such cosmic rays, with kinetic
energies below the energy threshold for pion formation, $\lesssim
280$\,\unit{MeV}, are very efficient in ionising both atomic and molecular
hydrogen. Subsequently, ionised hydrogen triggers an ion-driven chemistry
network, resulting in the formation of molecular ions whose abundances are
directly linked with the ionisation rate. As a consequence, large abundances
of these molecules, detected via their characteristic rotation-vibration
lines, serve as an indicator of an enhanced ionisation rate compared to the
ionisation rate induced by the diffuse photon and cosmic ray fluxes in the
Galactic plane. In order for astrospheres and wind bubbles to be identified
as (sub)TeV cosmic ray acceleration sources, cosmic ray-induced ionisation
needs to dominate over photoionisation at these objects and \mbox{ionisation
signatures} in spatial correlation with the gamma rays must be found. A
correlation study of this kind was performed in \citet{schuppan14} for
supernova remnants associated with molecular clouds. The model developed
therein can be adapted to astrospheres and wind bubbles of massive stars.

\section{Cosmic ray-induced ionisation}

For a proper description of the propagation of low-energy ($E_\mathrm{p}<
280$\,\unit{MeV}) cosmic ray protons in the vicinity of a cosmic ray
accelerator with general momentum loss, the transport equation
\begin{align} \label{transp_eq_large}
\frac{\partial n_\mathrm{p} \left( \vec{r}, p, t \right)}{\partial t} - D(p)
\Delta n_\mathrm{p} \left(\vec{r}, p, t \right)
& - \frac{\partial}{\partial p} \Big(b(p) \cdot n_\mathrm{p} \left( \vec{r}, p, t \right) \Big) \nonumber \\
& = Q(\vec{r},p,t)
\end{align}
has to be solved, where $n_\mathrm{p} \left(\vec{r}, p, t \right)$ is the
differential number density of cosmic ray protons, $D(p)$ is the scalar
diffusion coefficient, $\Delta$ denotes the Laplace operator, $b(p)$ is the
momentum loss rate, and $Q(\vec{r},p,t)$ is a source function which is
adapted to the astrophysical object under consideration. The low energy of
the cosmic ray protons allows for a scalar diffusion coefficient because then
there is no preferred \mbox{propagation direction,} whereas in a scenario
with highly relativistic motion, an anisotropic diffusion coefficient is
required. In \citet{schuppan14}, this transport equation is solved
analytically for an arbitrary source function. The Green's function $G \left(
\vec{r}, \vec{r}_0, p, p_0, t \right)$ of Eq.~(\ref{transp_eq_large}), i.e.,
the fundamental solution for a source term consisting of Dirac distributions
\begin{align*}
Q_G \left( \vec{r}, p, t \right) = \delta^3 \left( \vec{r} - \vec{r}_0
\right) \delta \left( p - p_0 \right)\delta(t),
\end{align*}
is given by
\begin{align} \label{eq:G_sol}
G \left( \vec{r}, \vec{r}_0, p, p_0, t \right) = &\frac{\Theta \left( p_0 - p
\right) \delta\left(t + \int_{p_0}^p {b(p')^{-1}\,\mathrm{d}p'}\right)}
{b(p) \cdot \left(4 \pi \int_{p}^{p_0} {D(p')/b(p')\, \mathrm{d}p'}\right)^{3/2}} \nonumber \\
& \cdot\exp \left(\frac{-\left(\vec{r} - \vec{r}_0 \right)^2}{4
\int_{p}^{p_0}{D(p')/b(p')\, \mathrm{d}p'}} \right),
\end{align}
where $\Theta(\cdot)$ is the Heaviside step function. This solution is
convoluted with the source function $Q(\vec{r}_0,p_0,t_0)$ of an astrosphere
or a wind bubble, yielding the differential number density of cosmic ray
protons at any position $\vec{r}$ at time $t\ge 0$ with momenta lower than
the injection momentum $p\le p_0$
\begin{align} \label{convol_G_Q}
&n_\mathrm{p} \left( \vec{r}, p, t \right) \nonumber \\
& = \iiint{G \left( \vec{r}, \vec{r}_0, p, p_0, t \right) Q\left( \vec{r}_0,
p_0, t_0 \right)\, \mathrm{d}t_0\,\mathrm{d}^3r_0\, \mathrm{d}p_0}.
\end{align}
The momentum-dependent component of the source function
$Q(\vec{r}_0,p_0,t_0)$ can be obtained from observational gamma ray data for
a hadronic formation scenario of the gamma rays \citep{schuppan12}, while the
spatial dependence is constructed reflecting the geometry of the source
region, e.g., a spherical (shell) volume with a constant emission over a
specific period of time.

The ionisation rate of molecular hydrogen induced by low-energy cosmic ray
protons can be calculated following \citet{padovani09}
\begin{align} \label{ion-pado}
\zeta^{\chem{H_2}} \left( \vec{r}, t \right) = \left(1 + \phi \right)
\int_{E_{\min}}^{E_{\max}} \frac{\mathrm{d}^3 N_\mathrm{p} \left( \vec{r},
E_\mathrm{p}, t \right)}{\mathrm{d}E_\mathrm{p} \,\mathrm{d}A\,\mathrm{d}t}~
\sigma_{\mathrm{ion}}^{\chem{H_2}} \left( E_\mathrm{p} \right) \,
\mathrm{d}E_{\mathrm{p}}\,,
\end{align}
where $\mathrm{d}^3 N_\mathrm{p}/\left(\mathrm{d}E_\mathrm{p} \,\mathrm{d}A\,
\mathrm{d}t \right)$ is the differential flux of cosmic ray protons,
$\sigma_{\mathrm{ion}}^{\chem{H_2}} \left( E_\mathrm{p} \right)$ is the
direct ionisation cross-section of molecular hydrogen, and $\phi\approx 0.6$
denotes the number of secondary ionisation events per primary ionisation
\citep{cravens78}. The differential flux can be derived from the differential
cosmic ray proton number density (Eq.~\ref{convol_G_Q}) by differentiating
with respect to the cosmic ray particle kinetic energy $E_\mathrm{p}$ and
subsequent multiplication with the effective cosmic ray particle speed, which
is given by a superposition of diffusion and the relativistic, kinematic
motion. This total speed is also used to link the space and time components,
leading to a description which assumes all particle speeds to be equal to the
statistic expectation value. Then, the time since the injection, which is
usually unknown, is not required.

In this model, the effects of magnetic fields, particularly magnetic
focussing and magnetic mirroring, are not accounted for. Taking them into
consideration, \citet{padovani2011} found that in dense molecular clouds
magnetic fields lead to a net decrease of the ionisation rate induced by
cosmic rays by a factor of, on average, 3--4. In less dense regions, like
swept up stellar ejecta at the shocks of astrospheres or stellar wind
bubbles, the decrease of the cosmic ray-induced ionisation rate is lower.

\section{Cosmic ray composition}

In \citet{padovani09}, a method taking not just ionisation induced by cosmic
ray protons, but also by heavier nuclei into account was introduced using the
Bethe--Bloch approximation for the direct ionisation cross-section
\citep{bethe33} and a correction factor for the differential cosmic ray flux
in Eq.~(\ref{ion-pado}),
\begin{align*}
\eta = \sum_{k \ge 2}{Z_k^2} \cdot \frac{f_k}{f_\mathrm{p}},
\end{align*}
where $k$ is the atomic number of the cosmic ray particles heavier than
hydrogen with the corresponding charge number $Z_k$ and $f_k/f_\mathrm{p}$ is
the relative abundance of cosmic ray particles of atomic number $k$ with
respect to cosmic ray protons. For relative abundances detected in the solar
system \citep{anders89,meyer98}, the correction was calculated in
\citet{padovani09} and \citet{indriolo09} as $\eta_{\odot}\approx 0.5$. At
the acceleration regions, the composition may well differ. In order to obtain
data with as little influence from the Sun as possible, observational data of
the cosmic ray composition taken during solar minima, when the impact of the
Sun on the cosmic ray flux is minimal, should be used. Therefore, the
calculation of the correction is done for observational data from the solar
minima in 1977 \citep{simpson83} or in 1988 \citep{meyer98}. The resulting
corrections are $\eta_{77}\approx 1.8$ and $\eta_{88}\approx 1.4$ (in
agreement with the results in \citealp{indriolo09}), respectively, indicating
that, on the one hand, the impact of the Sun on the composition of the cosmic
ray spectrum at low energies cannot be neglected and, on the other hand, that
the composition of the cosmic ray spectrum at the acceleration regions shows
significantly larger abundances of nuclei with $Z>2$ than the corresponding
abundances in the solar system. Adapting the momentum loss $b(p)$, which for
cosmic ray protons is dominated by Coulomb losses and adiabatic deceleration
in the relevant kinetic energy range of $10\,\unit{MeV}\le E_\mathrm{p}\le
280\,\unit{MeV}$ \citep[see, e.g.,][]{lerche-schlick1982} to cosmic ray
particles with $k\ge 2$, the solution given in Eq.~(\ref{eq:G_sol}) can also
be used to describe the propagation of those heavier cosmic ray particles,
resulting in an even more accurate calculation of the cosmic ray-induced
ionisation rate.

\section{Photoionisation}

Besides cosmic ray-induced ionisation, photoionisation is the other process
capable of efficient ionisation of ambient matter with soft X-rays and UV
radiation being particularly efficient. Consequently, photoionisation needs
to be exceeded by ionisation induced by hadronic low-energy cosmic rays in
order to attribute enhanced ionisation rates to those cosmic rays via the
suggested correlation study for astrospheres or stellar wind bubbles with
nearby clouds acting as a target for both photons and cosmic rays. Therefore,
spatially resolved spectral measurements of photon fluxes are important for
estimates of the photoionisation rate. The attenuation of photon fluxes due
to traversed matter can be described by means of the Beer--Lambert law
\begin{align} \label{B-L}
F_{\gamma} \left( \vec{r}, E_\gamma \right) = F_{\gamma, 0} \left( E_\gamma
\right) \cdot\exp{\big( -\tau \left( \vec{r}, E_\gamma \right)\big)} ,
\end{align}
where $F_{\gamma}$ is the differential photon flux, $F_{\gamma,0}$ is the differential photon flux at
the surface of a cloud located near the source, and $\tau(\vec{r},E_\gamma)$ is the
optical depth inside the cloud.
For the special case of matter distributed homogeneously with density $n_{\rm H}$,
the optical depth can be expressed in terms of the distance from the source region as
\begin{align}
\tau \left( \vec{r}, E_\gamma \right)_\mathrm{hom} = | \vec{r} | \cdot
n_\mathrm{H} \cdot \sigma_\mathrm{pa} \left(E_\gamma \right),
\end{align}
where $\sigma_\mathrm{pa} \left(E_\gamma \right)$ is the total
photoabsorption cross-section. This formula leads to an overestimate of the
photon flux at large distances from the source region, because it does not
account for scattering, which would in effect increase the distance the
photons travel. Hence, the energy attenuation of the photon flux within the
cloud is underestimated. The photoionisation rate of molecular hydrogen can
be calculated following \citet{maloney1996}
\begin{align} \label{photoion}
\zeta^{\chem{H_2}}_{\gamma} \left( \vec{r} \right) =
\frac{f_\mathrm{i}}{I_{\chem{H_2}}}\int_{E_{\min}}^{E_{\max}} F_{\gamma}
\left( \vec{r}, E_\gamma \right)\cdot E_\gamma \cdot \sigma_\mathrm{pa}
\left(E_\gamma \right)\,\mathrm{d}E_\gamma,
\end{align}
where $f_\mathrm{i}\approx 0.4$ is the fraction of the photon energy absorbed
by ambient matter leading to ionisation \citep{voit1991,dalgarno1999}, and
$I_{\chem{H_2}}=15.4$\,\unit{eV} is the ionisation potential of molecular
hydrogen. The integral describes the total energy deposition by the absorbed
photons per hydrogen nucleus. A parametrised description of the total
photoabsorption cross-section $\sigma_\mathrm{pa}(E_\gamma)$ is provided as
an empirical, broken power-law fit to experimental data.

The main issue computing the photoionisation rate is obtaining a solid
estimate of the photon flux at low energies, i.e., $E_\gamma\lesssim
0.1$\,\unit{keV}. Since the total photoabsorption cross-section increases
toward lower photon energies, the corresponding photons can contribute
significantly to the photoionisation rate, depending on spectral shape of the
photon flux in this energy range. Hence, observing the low-energy photon
fluxes is an important task in estimating the photoinduced ionisation.
Particularly the UV component of the photon flux is very efficient in
ionising ambient hydrogen close to the surface of a cloud and, thus, has a
great impact on the photoionisation rate.

In contrast to the examination of, e.g., supernova remnants associated with
molecular clouds, a solid estimate of the low-energy photon fluxes in
astrospheres can be provided much more reliably via the classification of the
star and usage of the typical corresponding black-body radiation spectrum.
Since in most situations there is only very little information about the UV
domain of the photon flux, which is a major issue for the estimate of
photoionisation rates, astrospheres are promising targets for the suggested
correlation study in this regard.

\section{Ionisation signatures}

The total ionisation rate, i.e., the sum of the photoinduced and the cosmic
ray-induced ionisation rates, can be used as input for astrochemistry models
in order to compute signatures indicating an enhanced ionisation rate, e.g.,
in the form of rotation-vibration lines of molecular ions formed by
ion-driven chemistry, detectable with instruments like ALMA. While the
ionisation rate is not directly detectable, its impact on ambient matter,
e.g., diffuse molecular clouds in star-forming regions containing wind
bubbles of massive stars, or interstellar matter swept up by stellar winds,
results in the formation of molecular ions that are otherwise not formed in
observable abundances. Among the molecular ions formed in these environments,
\chem{H_3^+} is especially suitable as a tracer of ionisation, because its
formation scheme is rather simple and directly linked to the ionisation rate
of hydrogen \citep[see, e.g.,][]{indriolo10}.

An example for a tool to calculate rotation-vibration lines of molecular ions
is the radiative transfer code RADEX \citep{Radex2007}. It allows rapid
computation of, e.g., rotation-vibration emission line intensities of
molecules, including \chem{H_3^+}, as was shown in \citet{becker11} for
molecular clouds near supernova remnants, based on input parameters such as
the abundances of several atomic and molecular species, the electron number
density, and a few others. Furthermore, it provides different geometries and
the option to include a background radiation field.

Some of the molecules and molecular ions are best seen in absorption, so
their rotation-vibration lines need to be detected as absorption features in
the spectra of background sources. While for supernova remnants associated
with molecular clouds it is difficult to find an adequate (sufficiently
bright) background source with known photon spectrum, this does not pose a
problem for astrospheres, since the classification of the star provides a
reasonable estimate of the local radiation field at the corresponding
wavelengths in the IR- and submm-domain.

Another method to determine the ionisation rate via observations is to look
for the ratio of the abundances of certain molecular species, e.g., the
\chem{DCO^+}$/$\,\chem{HCO^+} ratio, but any other ratio of molecular
abundances sensitive to the ionisation rate can be chosen as well. A
detection of either these ratios or predicted rotation-vibration lines,
whether in emission or absorption, would be a strong hint at an enhanced
ionisation rate.

In the supernova remnant complexes IC~443, W28, and W51C, enhanced ionisation
rates were observed by \citet{indriolo10}, \citet{nicholas2011}, and
\citet{ceccarelli2011}, respectively, via \chem{H_3^+} abundances, ammonia
(\chem{NH_3}) emission lines, and the \chem{DCO^+}$/$\,\chem{HCO^+} ratio,
respectively, indicating that the detection of enhanced ionisation rates can
be performed with current-generation telescopes. With ALMA, even more
detailed studies of the substructure of interesting objects are possible.

\conclusions A correlation study of both gamma rays formed via proton-proton
interactions of cosmic ray protons of kinetic energies $E_\mathrm{p}\ge
280$\,\unit{MeV} with ambient matter and ionisation signatures induced by
cosmic ray protons of lower energies can be used to examine cosmic ray
acceleration in many different astrophysical sources, such as supernova
remnants, astrospheres, and wind bubbles of massive stars. To this end, the
position-dependent ionisation profiles induced by photons and cosmic ray
protons can be calculated in order to determine the total ionisation rate,
which in turn leads to the formation of molecular ions. These ions can be
observed via their rotation-vibration line spectra with current-generation
telescopes, among which ALMA is particularly well-suited due to its high
spatial resolution, while the detection of the gamma rays can be done with
instruments such as FermiLAT, H.E.S.S., and CTA. Observational evidence of
enhanced ionisation rates which can only be explained by cosmic rays can help
establish astrospheres and wind bubbles of massive stars as sources of
(sub)TeV cosmic rays, leading to a better understanding of these objects.

\begin{acknowledgements}
We would like to thank John H.~Black, Reinhard Schlickeiser, Marco Padovani,
and Nick Indriolo for helpful and inspiring discussions. We are also grateful
to the anonymous referee whose comments significantly improved the
manuscript.\hack{\newline}
 \hack{\newline}
 Edited by: K.~Scherer \hack{\newline}
Reviewed by: M.~Padovani and one anonymous referee
\end{acknowledgements}

\end{document}